\documentclass[aps,prb,reprint,showpacs,amsmath,amssymb,amsfonts,floatfix]{revtex4-1}

\usepackage{bm}
\usepackage{graphicx}
\usepackage[bookmarks,bookmarksopen,bookmarksnumbered,colorlinks,linkcolor=red,linktocpage,citecolor=blue,urlcolor=cyan, pdfpagemode=UseOutline]{hyperref}
\usepackage{epstopdf}

\def\bea{\begin{eqnarray}}
\def\eea{\end{eqnarray}}
\def\ben{\begin{equation}}
\def\een{\end{equation}}
\def\benu{\begin{enumerate}}
\def\enu{\end{enumerate}}

\def\bei{\begin{itemize}}
\def\eei{\end{itemize}}
\def\benu{\begin{enumerate}}
\def\enu{\end{enumerate}}

\def\sss{\scriptscriptstyle\rm}

\def\1var{(\bx_1...\bx\N)}

\def\br{{\bf r}}

\def\bx{{x}}

\def\xc{_{\sss XC}}

\def\Hxc{_{\sss HXC}}

\def\N{_{\sss N}}

\def\sph_int{ {\int d^3 r}}

\newcommand{\vect}[1]{\mathbf{#1}}
\newcommand{\matelem}[3]{\left\langle #1 \left| #2 \right| #3 \right\rangle}

\providecommand{\abs}[1]{\left|#1\right|}

\newcommand{\parref}[1]{(\ref{#1})}

\begin{document}
\title{Application of object-oriented programming in a time-dependent density-functional theory calculation of exciton binding energies}
\author{Zeng-hui Yang and Carsten A. Ullrich}
\affiliation{Department of Physics and Astronomy, University of Missouri, Columbia, MO 65211, USA}
\date{\today}

\begin{abstract}
This paper discusses the benefits of object-oriented
programming to scientific computing, using our recent
calculations of exciton binding energies with time-dependent
density-functional theory (arXiv: 1302.6972) as a case study. We
find that an object-oriented approach greatly facilitates the
development, the debugging, and the future extension of the
code by promoting code reusing.  We show that parallelism is
added easily in our code in a object-oriented fashion with
ScaLAPACK, Boost::MPI and OpenMP.
\end{abstract}

\maketitle

\section{Introduction}
\label{sect:intro}

\subsection{Object-oriented programming in scientific computing}

A plethora of programming languages are available nowadays.
While all are of academic interest, at least in principle, only
a few are used in large scale computations, and these can be
roughly categorized into two groups: those promoting imperative
programming (such as Fortran and C) and those promoting
object-oriented programming (OOP). For the usual implementations of imperative programming,
each statement of the programming language can be translated
into simple machine code constructs; while OOP is more abstract by considering data as
opaque objects, each having a unique set of methods to
manipulate data. Only a pre-defined set
of operations are allowed on data, and the internal representation of data
is not visible to the outside; this indirectness has been
proven to be very helpful to the designing, the writing, and
the modifying of the program. OOP was not widely used until the
emergence of programming languages that support expressing OOP
concepts natively, most notably C++.\cite{S13,CppPrimer} C++
provides both the programming convenience of OOP (and other higher-level
paradigms) and, most importantly, the runtime efficiency of C.
C++ makes OOP not only of academic interest but also of
practical use since the runtime efficiency is not
compromised much.

To better distinguish these two paradigms, the concept of
`coupling'\cite{DesignPatterns,EffectiveCpp,MoreEffectiveCpp}
need to be introduced. If changing a certain part of the
program induces the need to change another part, these two
parts of the program are coupled. Since imperative programming
is closer to machine code, one often manipulates raw data
directly. This is inherently fast, but this also makes the
program dependent on how the raw data is represented, creating
tighter coupling between different parts of the program. If the
representation of the data is to be changed later, a large
amount of the code need to be changed. On the other hand, OOP
promotes the separation of interface (what manipulation is to
be done on the data) from implementation (how the data is
actually manipulated). As a consequence, good OOP design
achieves loose coupling and greater ease of maintenance.

The use of OOP in quantum mechanics was pioneered in the work
of DFT++ by Ismail-Beigi,\cite{IA00} and since then many other
applications have emerged, such as S/PHI/nX,\cite{BFDI11}
JDFTx,\cite{SLA12} and others. Despite the vast advantage
provided by OOP, a considerable number of programs used in
computational physics are still written with imperative
programming in mind. While the runtime efficiency is good, the
codes are harder to learn, to modify, and to extend. We have
recently calculated exciton binding energies with
time-dependent density-functional theory (TDDFT),\cite{YU13}
and we have developed a C++ code\cite{supplemental} for this
calculation employing OOP. We provide a case study of our code
to demonstrate how OOP is helpful in computational physics, and
we thus hope to promote its usage.

\subsection{Background of the physical problem}
\label{sect:intro:bkgnd}

We briefly introduce the physical problem here to provide a
context for our code, while the computational details of the
problem are described later. Excitons are coupled electron-hole
pairs in solids,\cite{HK09,U12} and they show up in optical
absorption spectra as discrete absorption peaks below the band
gap. Many-body theories such as the Bethe-Salpeter equation
(BSE)\cite{HS80,ORR02} describe excitons well, but the
computation is costly, and thus its use is limited. An
alternative to the BSE is TDDFT,\cite{RG84,U12} which is a
promising excited-state electronic structure method widely used
for finite systems such as molecules, but is gaining popularity
for periodic systems. Density functional methods balance
accuracy and computational cost by calculating an auxiliary
non-interacting Kohn-Sham system that has the same electronic
density as the real, interacting system. Though TDDFT is
formally exact, one needs to approximate the
exchange-correlation (xc) many-body effects in practice;
numerous approximations for the xc potential $v\xc$ and for the
xc kernel $f\xc = \delta v\xc/\delta n$ ($n$ is the
one-particle density) have been successfully applied to
describe  electronic structure and excitations in materials.

The excitons have been difficult to obtain from a TDDFT
calculation. While the main difficulty is an unusually
stringent requirement on the xc kernel,\cite{GGG97} the
commonly used calculation approach is also not suitable for
exciton binding energies.\cite{YLU12} We have developed an
alternative approach for calculating exciton binding energies
in TDDFT in our recent work.\cite{YU13} Our code constructs and
solves an eigenvalue problem as follows (atomic units
$e=\hbar=m_e=1/4\pi\epsilon_0=1$ are used throughout the paper
unless otherwise mentioned):
\begin{equation}
\sum_{(mn)}\left[\delta_{im}\delta_{jn}(\epsilon_j-\epsilon_i)+F\Hxc^{(ij)(mn)}\right]\rho^{(mn)}(\omega)=\omega\rho^{(ij)}(\omega),
\label{eqn:intro:TDDFTworking}
\end{equation}
where $\epsilon$ are the Kohn-Sham orbital energies of the
underlying electronic ground-state calculated beforehand from
the ABINIT code,\cite{ABINIT} $i$, $m$ denote valence bands,
$j$, $n$ denote conduction bands, the eigenvalue $\omega$ is
the excitation frequency, and the corresponding eigenvector
$\rho$ describes how single-particle excitations combine into
the real excitation. The Hartree-exchange-correlation (Hxc)
matrix $F\Hxc=2F_\text{H}+F\xc$ is composed of the Hartree part
$F_\text{H}$ and the xc part $F\xc$, defined as
\begin{eqnarray}
F_\text{H}^{(ij\vect{k})(mn\vect{k}')}&=&\frac{1}{V}\sum_{\vect{G}\ne0}\frac{4\pi}{\abs{\vect{G}}^2}
\matelem{j\vect{k}}{e^{i\vect{G}\cdot\br}}{i\vect{k}}
\nonumber\\
&&\times
\matelem{m\vect{k}'}{e^{-i\vect{G}\cdot\br}}{n\vect{k}'}
\label{eqn:intro:fHworking}
\end{eqnarray}
and
\begin{multline}
F\xc^{(ij\vect{k})(mn\vect{k}')}=\frac{2}{V}\sum_{\vect{G}\vect{G}'}f_{\text{xc},\vect{G}\vect{G}'}(\vect{q}=0)\\
\times\matelem{j\vect{k}}{e^{i\vect{G}\cdot\br}}{i\vect{k}}\matelem{m\vect{k}'}{e^{-i\vect{G}'\cdot\br}}{n\vect{k}'}.
\label{eqn:intro:fxcworking}
\end{multline}
Here, $V$ is the volume of the crystal, $\vect{G}$ and
$\vect{G}'$ are the reciprocal lattice vectors, $\vect{k}$ is
the wavevector in the first Brillouin zone, and $\vect{k}$
together with band index $i$, $j$, $m$, or $n$ specifies the
Kohn-Sham orbitals.  The approximation for the xc kernel $f\xc$
is chosen by the user for each calculation.

\begin{figure*}[t]
\includegraphics[width=\textwidth]{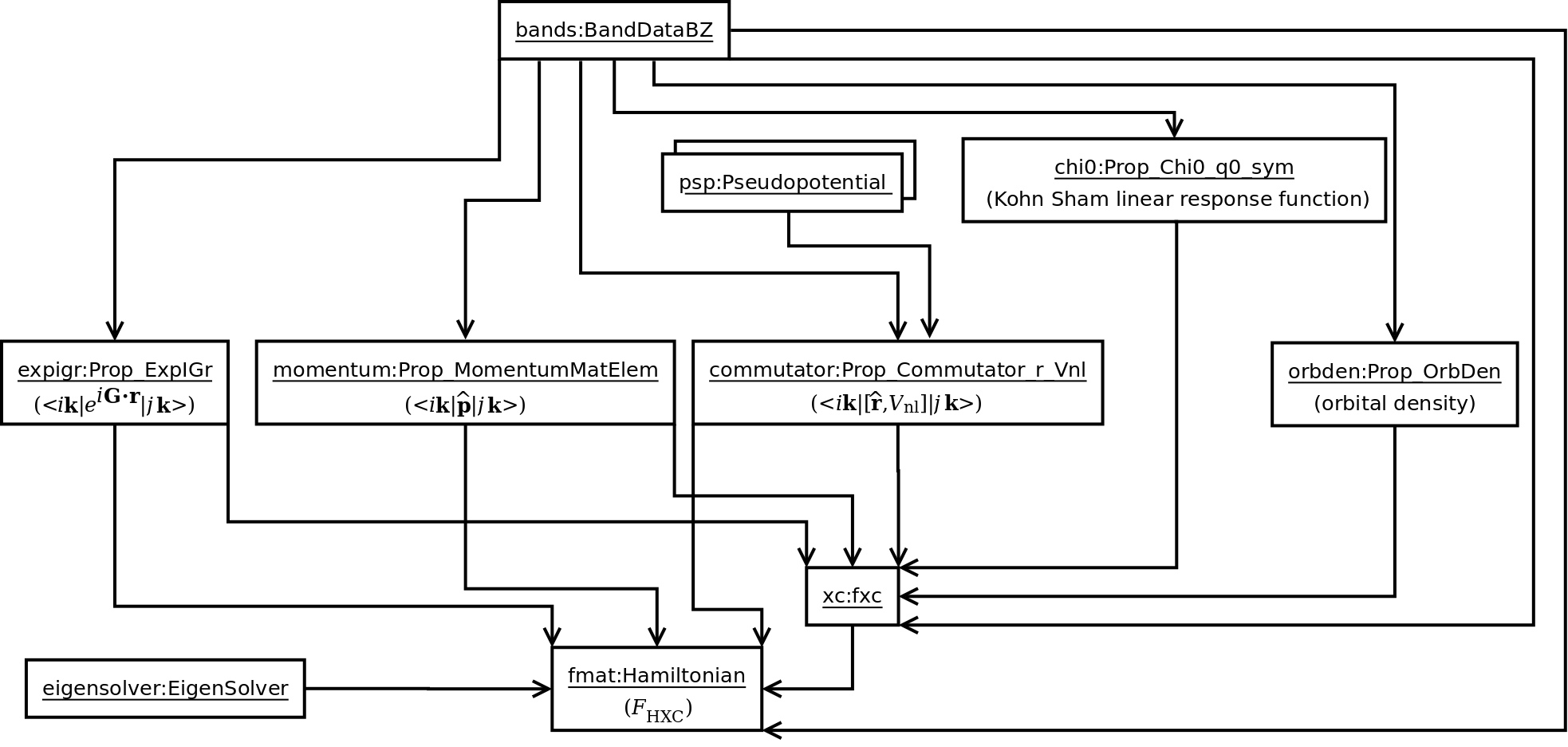}
\caption{UML\cite{UML} collaboration diagram of our code, where
every node is an object. Only the most important objects are
shown. The arrows denote dependence between objects, with the
arrowhead side depending on the arrowtail side. Only those
objects which are directly related to the calculation are
shown.} \label{fig:design:flowchart}
\end{figure*}

The formalism above assumes spin-unpolarized systems and does
not explicitly treat spin. It cannot describe spin-flip
triplet excitations. Proper treatment of spin-flip excitations
requires the non-collinear spin formalism.\cite{U12} For
spin-unpolarized systems, however, there is a shortcut for
calculating triplet excitations. We define
\begin{equation}
f\xc^\text{singlet}=\frac{f\xc^{\uparrow\uparrow}+f\xc^{\uparrow\downarrow}}{2},\quad
f\xc^\text{triplet}=\frac{f\xc^{\uparrow\uparrow}-f\xc^{\uparrow\downarrow}}{2},
\label{eqn:intro:spinfxc}
\end{equation}
where $f\xc^{\uparrow\uparrow}$ and $f\xc^{\uparrow\downarrow}$
are parts of the spin-dependent xc kernel,\cite{U12} and they
need to be approximated as well in practice. Solving Eq.
\parref{eqn:intro:TDDFTworking} with $f\xc^\text{singlet}$ and
$f\xc^\text{triplet}$ yields singlet and triplet excitations,
respectively.

The $\vect{G}=\vect{G}'=0$ part (so-called `head') of $f\xc$
requires special treatment, since it diverges as $q^{-2}$ as
$q\to0$. For $q\ne0$, the matrix element
$\matelem{j\vect{k}}{e^{i\vect{G}\cdot\br}}{i\vect{k}}$ in
Eq. \parref{eqn:intro:fHworking} and \parref{eqn:intro:fxcworking} becomes
$\matelem{j\vect{k}}{e^{i(\vect{q}+\vect{G})\cdot\vect{r}}}{i\vect{k-q}}$,
and in the limit of $q\to0$ the divergence in $f\xc$ is
canceled out. The head contribution to Eq.
\parref{eqn:intro:fxcworking} is then calculated as
\begin{equation}
\begin{split}
&\lim_{\vect{q}\to0}\matelem{j\vect{k}}{e^{i(\vect{q}+\vect{G})\cdot\vect{r}}}{i\vect{k}-\vect{q}}
\matelem{m\vect{k}'-\vect{q}}{e^{-i(\vect{q}+\vect{G})\cdot\vect{r}}}{n\vect{k}'}\\
&\quad\quad \times\frac{2}{V}f_{\text{xc},00}(\vect{q})\\
&=\frac{\matelem{j\vect{k}}{\hat{\vect{p}}-i[\hat{\vect{r}},V_\text{nl}]}{i\vect{k}}}{\epsilon_{j\vect{k}}-\epsilon_{i\vect{k}}}
\frac{\matelem{m\vect{k}'}{\hat{\vect{p}}-i[\hat{\vect{r}},V_\text{nl}]}{n\vect{k}'}}{\epsilon_{n\vect{k}'}-\epsilon_{m\vect{k}'}}\\
&\quad\quad \times\lim_{\vect{q}\to0}\frac{2q^2}{V}f_{\text{xc},00}(\vect{q}),
\end{split}
\label{eqn:intro:headcontribution}
\end{equation}
where $\hat{\vect{p}}$ is the momentum operator,
$\hat{\vect{r}}$ is the position operator, and $V_\text{nl}$ is
the non-local part of the pseudopotential. In our
study,\cite{YU13} we are only interested in the exciton binding
energy instead of the entire optical spectrum, so we only need
to include a few bands in Eq.
\parref{eqn:intro:TDDFTworking}, which simplifies the problem.

This paper is structured as follows: in Sect. \ref{sect:design}
we examine the needs of the exciton project, and we discuss the
design of our code based on these needs; we then show the
implementation details in Sect. \ref{sect:implementation} to
demonstrate the role of OOP in the development of our code; we
conclude in Sect. \ref{sect:conclusion}.

Several pseudo-code examples are given in the text; the reader
is asked to refer to the supplemental material for the actual
codes. The calculation results were presented in Ref.
\onlinecite{YU13} and are not repeated here.

\section{Design}
\label{sect:design}

\subsection{More general remarks on OOP}


Before going into specifics of the design of our code, let us
discuss a few important features OOP. We mentioned in
Sect. \ref{sect:intro} that OOP views data as opaque objects
that are accessible only through a predefined set of methods,
through which the visible interface (methods) and the invisible implementation
(representation of data and their manipulation) are separated,
allowing one to program without worrying about details. An
object has a certain predefined type (known as `class' in C++)
that defines interface and encapsulates implementation, and one class
can have more than one object instances.

Another defining feature of OOP is inheritance and
polymorphism. Inheritance derives an object type from a base
type by copying the code of the base type into its own
definition. This achieves code reuse and uniformity, such
that a common operation can be defined within a base type, and
if later it is to be modified in the base type, the change
automatically applies to all the derived types; but more
importantly, inheritance models a logical connection between
types, most commonly an `is a' relation\cite{EffectiveCpp} [for
example, a long-range correction (LRC)
 xc kernel (derived type) \emph{is an} xc kernel (base type)].
This leads to the so-called polymorphism---an object of the
derived type can be used anywhere an object of the base type is
required, since the object of the derived type `is an' object
of the base type.

Polymorphism further decouples interface and implementation, so
that the base type provides a uniform interface of methods, and
the derived types do the actual calculation. For example,
calculation of the $F\xc^{(ij\vect{k})(mn\vect{k}')}$ matrix
elements in Eq.
\parref{eqn:intro:fxcworking} requires the value of the
reciprocal space xc kernel $f\xc$ (defined as a base type), but
this calculation does not need to know either the actual choice
of $f\xc$ or what types of $f\xc$ are available (which all
derive from the common $f\xc$ base type and do the calculation);
since this calculation does not rely on details of
$f\xc$ themselves, one can easily write a new derived type of
$f\xc$ and expect it to work like other $f\xc$'s, without
needing to change the code for calculating the
$F\xc^{(ij\vect{k})(mn\vect{k}')}$ matrix elements. In C++,
polymorphism is provided through virtual
methods.\cite{CppPrimer,S13}

\begin{figure*}[t]
\includegraphics[width=\textwidth]{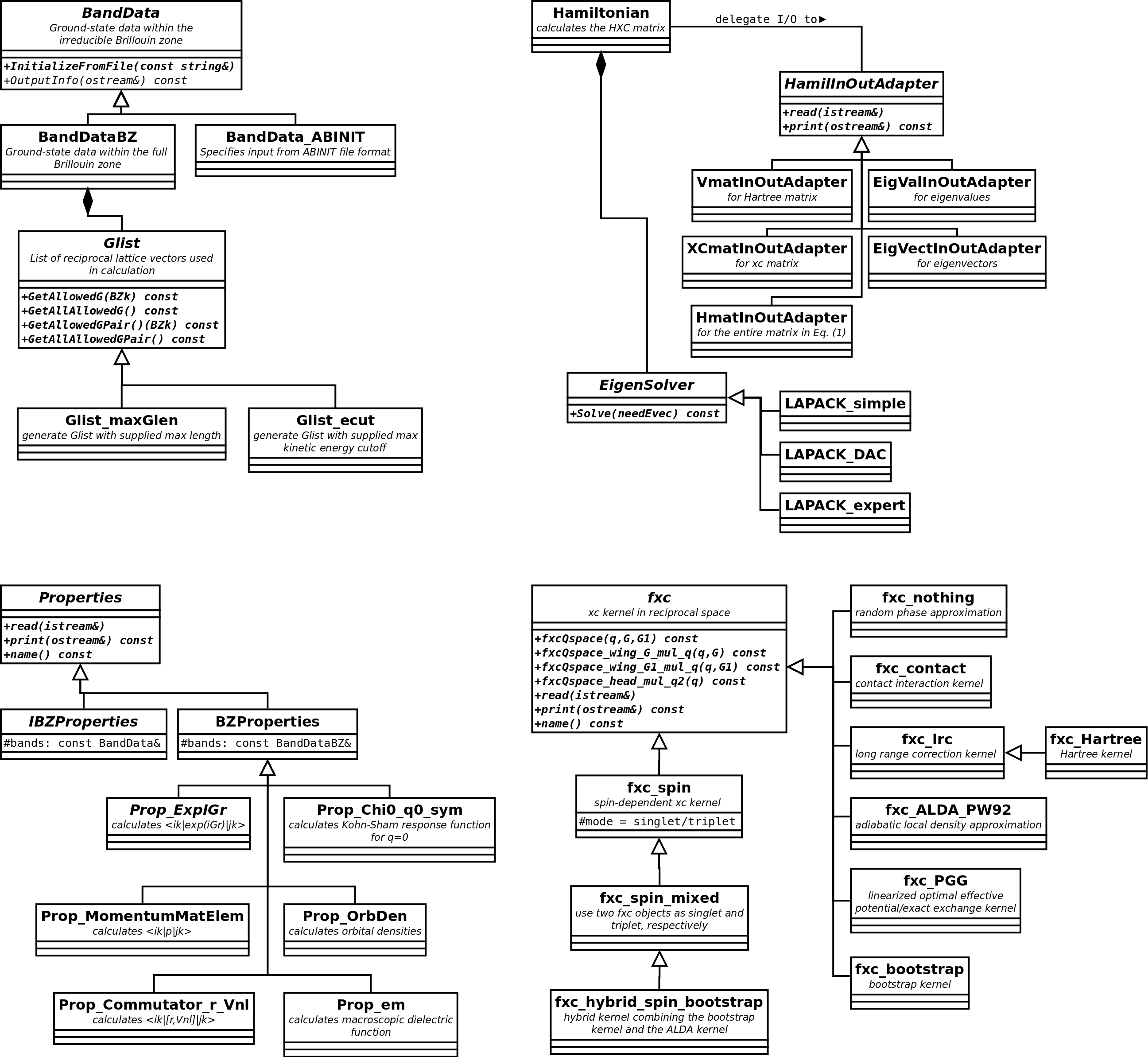}
\caption{UML\cite{UML} class diagram of the important classes in the
exciton project. Only the virtual methods of base classes and
important member variables are shown.}
\label{fig:design:hierarchy}
\end{figure*}

It should be noted that the main purpose of programming
paradigms such as OOP is to provide a framework with which
programmers can structure, modify, or extend their program with
ease, and thus increase the productivity and the development
speed. The runtime efficiency is not the primary concern of
programming paradigms. Object-oriented programs generally produce more
machine codes for the same task; object-oriented codes are
harder to optimize by the compiler due to no direct mapping to machine codes;
the memory access pattern
is more difficult to predict---all these issues make object-oriented
codes run slightly slower than programs written with imperative programming languages.
The OOP compensates the runtime
efficiency lost with a huge gain in development and maintenance
efficiency.

OOP alone does not guarantee good programming quality, and one
needs to design the object structure with care. We identify the
following characteristics of our exciton project\cite{YU13} that guide the
design of our code: this project is on a specific problem
instead of general quantum physics; this project has been and
will be worked on by different people; we do not want to
reinvent the wheel, and thus we need to use functionalities
provided by other codes as much as possible; we want to focus
on the physical problem instead of numerical details, so we use
specialized libraries; the focus of the project is on
developing the theory instead of merely on calculating numbers.

With these characteristics in mind, the general requirements on
the code are as follows: lightweight, fast to develop, easy to
access and to extend, adaptable to different libraries,
efficient to run moderate-sized calculations, and sufficient
runtime flexibility to allow for rapid testing of ideas. We
show the overall object layout in Fig.
\ref{fig:design:flowchart}.

OOP helps the structuring the program by ab\-strac\-tion---the
type of data is not determined by the representation, but by
the allowed methods manipulating it. How the data is
represented and how the methods change the data are
implementation details that should not have any influence on
the user of the data; thus, the user can avoid being
sidetracked by details. But a common pitfall in the practice of
OOP is to overly generalize the object concept and complicate
simple, unambiguous operations. Therefore it is important not
to lose sight of the goal of the code in order to determine the
level of abstraction and granularity of objects.

For example, the Hxc matrix $F\Hxc$ in Eq.
\parref{eqn:intro:TDDFTworking} contains a large portion of
data which are logically inseparable. The construction
of this matrix involves complicated calculations and must be
done efficiently, and the role of the matrix is pre-determined by Eq.
\parref{eqn:intro:TDDFTworking}. Packaging the involved
calculations inside an \texttt{fmat} object is a good
abstraction, since this decouples it from other important
operations the code needs to perform---such as reading the
input files, preparing ground-state orbitals and so on, so that
these operations can be changed with confidence that such
manipulations do not affect the calculation inside
\texttt{fmat}. On the other hand, to treat each matrix element
of $F\Hxc$ as an object would be unnecessarily complicated
for the use in our exciton study\cite{YU13}.

\subsection{Specifics of the project}

In this project, we quickly found that there was a need to
develop our own code. Although this is mainly because of the
calculation being a new approach for solids not available in
existing TDDFT codes, another important reason is that the
rigid, tightly-coupled structure of these codes deters
modification and extension. It was  surprising to see how many
changes are needed to the existing Fortran-based codes to
implement the calculation described in Sect.
\ref{sect:intro:bkgnd}. We have been using the ABINIT code to
calculate the ground-state data, so we use it as an example in
the following. However, the problem described here is general
with imperative programming, which is prevalent in existing
electronic structure codes.

For example, ABINIT stores the ground-state orbitals in
multi-dimensional arrays, and this poses three major problems.
First, the source code itself (aside from comments) does not tell the meaning of each
dimension, because as an imperative programming
language Fortran is close to machine code, and therefore lacks
self-documenting ability.
Second, subroutines like normalization, symmetry
operation, calculation of density and so on are logically
related to the manipulation of orbitals, but the source code does not have the
means for representing this relation. This information can only
be discovered by reading other parts of the code or the
documentation, during which more questions may emerge.
Also, the compiler is not able to check this logical connection to
ensure these operations are applied to valid data.
Third, all of the raw data are accessible at once, and this risks
introducing errors in the code accidentally. For example, suppose a subroutine
needs to change a certain part of the data, but a programming bug
affects other parts of the data as well; the compiler cannot detect
this bug since nothing in the program explicitly forbids it.
Such a bug can remain hidden for a long time.

As a consequence, one needs to have knowledge of a great
portion of the code to be able to maintain even a small part of
it. Though not necessarily a problem for the original
developers, this tightly-coupled structure makes the code hard
to learn and to modify for users with particular needs. In
comparison, with good OOP design, the data is accessed through
specialized methods with explicit meaning, the related
operations on the data become inseparable parts of the object
interface, and access privileges to the data are differentiated
for different parts of the program. These features not only
make the structure of the program clearer, but also make it
possible to perform more error checking at compile-time rather than
at runtime, since complex relationships between data can be represented
in a more explicit manner.

We take advantage of polymorphism to provide extensibility, and
we demonstrate the extensibility here by examples. The
inheritance hierarchy of our code is shown in Fig.
\ref{fig:design:hierarchy}. We reuse ground-state data from
other codes, so we package these data in the \texttt{BandData}
class. To avoid being locked into a certain ground-state code,
the base BandData class defines a virtual
\texttt{InitializeFromFile} method, and it relies on derived
classes to do the actual data input. We have defined a
\texttt{BandData\_ABINIT} to read the output of the ABINIT
code, which is a pseudopotential code; changing to other band
structure codes such as the full-potential ELK code\cite{ELK}
are possible in the future. This will only require deriving a
new class from \texttt{BandData} implementing
\texttt{InitializeFromFile}.

Programming paradigms like OOP facilitate the development of
the source code, and the efficiency of the compiled code is not
the primary concern of the paradigms. The indirectness provided
by OOP makes the structure of the program clearer, but it also
incurs runtime cost: not only extra machine instructions, but
also a potential decrease in cache-efficiency. To avoid such
cost, we use special libraries such as
ScaLAPACK\cite{ScaLAPACK} and FFTW\cite{FFTW05} for numerical
calculations. The interfaces to these libraries are more
suitable for imperative programming, and we package the actual
calls to these libraries inside adapter objects so that we can
still benefit from OOP. This also provides the possibility to
change to other libraries if needed---the adapter defines a
uniform interface, and we only need to provide an
implementation of this interface using the desired library,
without worrying about the other parts of the code.

The main purpose of our code is to help our ongoing theoretical
studies of TDDFT. Unlike general-purpose codes that perform
established calculations, the type of calculations that our
code performs need to change from time to time. We provide the
runtime flexibility in several ways. Instead of hard-coding
specific pieces of information, we make our code read as much
input as possible at runtime to avoid re-compiling and to
achieve data reuse. Making many functions depend on user input
also calls for structured exception handling to detect human
error and to allow for possible recovery from errors.

For example, we have calculated the
exciton binding energies with different $f\xc$ in Eq.
\parref{eqn:intro:fxcworking} for the same material.
Some data like the matrix element
$\matelem{j\vect{k}}{e^{i\vect{G}\cdot\vect{r}}}{i\vect{k}}$ do
not change between such calculations, and having to
re-calculate it every time would be wasteful. In most cases we
only need the exciton binding energy as the output, but
occasionally we also need to calculate the entire spectrum from
the eigenvectors of Eq. \parref{eqn:intro:TDDFTworking}. The
spectrum does not need to be calculated every time, and it is
useful to allow its separate calculation after a calculation of
only the binding energy, without having to diagonalize the
$F\Hxc$ matrix in Eq.
\parref{eqn:intro:TDDFTworking} again.

We solve these problems by exporting the input/output (I/O)
option of each object to the user and by employing the C++
exception handling mechanism. The output that the user requires
determines what calculations are carried out, so that no extra
calculation would be necessary; when a specific object is
required during the calculation, the program searches for the
user-specified input data file for this object and attempts to
reuse the data; if the data file is corrupt or non-existent,
an C++ exception is thrown, and the exception handler in
the main program allows for recovery by calculating this
part of data from scratch. The detailed implementation will be
discussed below in Sect. \ref{sect:implementation}.

\section{Implementation}
\label{sect:implementation}

The calculation performed by our program is shown in Fig.
\ref{fig:design:flowchart}. The program starts by reading the
the ground-state data in the irreducible Brillouin zone
and pseudopotentials (\texttt{psp}); the ground-state data is then extended to
the full Brillouin zone by symmetry operations and stored in
\texttt{bands}. Properties required for later calculation are
then calculated beforehand in parallel. \texttt{expigr},
\texttt{momentum}, \texttt{commutator} are always calculated
since they are required in the calculation of Eq.
\parref{eqn:intro:fxcworking} and Eq.
\parref{eqn:intro:headcontribution}; other properties are only
calculated when the specified \texttt{fxc} needs them. The
calculation of the xc kernel $f\xc$ is then carried out in
\texttt{xc} with required properties and the ground state data.
Then, \texttt{fmat} calculates the matrices in Eq.
\parref{eqn:intro:fHworking} (V matrix) and Eq.
\parref{eqn:intro:fxcworking} (XC matrix), combines them into
the left-hand side matrix in Eq.
\parref{eqn:intro:TDDFTworking} (H matrix), and uses the provided
\texttt{eigensolver} to solve Eq.
\parref{eqn:intro:TDDFTworking}. In the following, we describe
certain details of the implementation of our code in order to
demonstrate the benefits of OOP.

\subsection{\texttt{BandData}}
The \texttt{BandData} class holds all the ground-state data
that is obtained from other codes. The class interface does not
expose any raw data, but it contains methods for getting the
values of the required data, such as the lattice vectors (in
real or reciprocal space), the volume of the unit cell, the
range of bands included in calculation, the orbital energies,
the Bloch functions and so on. This not only disallows outside
code to change the data, but also allows these methods to
perform additional operations aside from getting the data.

For example, after the first executable version of our code was
finished, we implemented an additional scissor correction
within the method for getting band energies to shift the
conduction band energies by a certain value. Other parts of the
code that use band energies continue to use this method as in
the previous version and require no changes. By contrast, one
can create a new subroutine for this correction in imperative
programming, but one is then forced to change the other parts
of the code to ensure compatibility, which is both tedious and
error-prone. One can also create a subroutine for getting the
band energies from the beginning, and implement this correction
within this subroutine; but this not only requires planning
long before its actual use, but also requires the programmer to
be aware of this subroutine and actively use it instead of
using the raw data. Since imperative programming cannot
represent the relation between this subroutine and the data,
the burden of knowing the detailed structure of the program is
on the programmer.

To decouple the structure of our code from any specific
ground-state code, the \texttt{BandData} class (Fig.
\ref{fig:design:hierarchy}) does not implement the reading of
actual data from provided files, but it defines the abstract
\texttt{InitializeFromFile} method and relies on derived
classes to implement the method. In our recent
study,\cite{YU13} we have used the ABINIT code\cite{ABINIT} for the
ground-state calculation, and we derive the class
\texttt{BandData\_ABINIT}, which implements
\texttt{InitializeFromFile} to read the data files produced by
ABINIT. We plan to use the ELK code\cite{ELK} in the future. Although ELK
uses a different basis set (linearized augmented plane waves
instead of ordinary plane waves), we can hide these details in
the new derived class which converts the basis set, so that
other parts of the code do not need to know how the raw data is
represented.

The ground-state band-structure codes usually use symmetry of
the system to simplify the calculation; in our case the
ground-state data only contains the $\vect{k}$-points in the
symmetry-reduced (irreducible) Brillouin zone. The TDDFT
calculation Eq.
\parref{eqn:intro:TDDFTworking} needs the entire Brillouin
zone, however. Also, different parts of the program may need to
work with different numbers of bands or $k$-points: the
\texttt{fmat} object may use fewer bands than the \texttt{xc}
object for faster calculation. We derive the
\texttt{BandDataBZ} class from \texttt{BandData} for two purposes: to generate
symmetry-related data and to make a selected part of the band
viewable.

For the first purpose, \texttt{BandDataBZ} is mostly an
extension to the \texttt{BandData} interface, providing more
methods to access the symmetry-related data, and inheritance
allows the code of general methods (such as getting the lattice
vector) to be reused. For the second purpose, the methods for
getting the band range and number of $\vect{k}$-points defined
in \texttt{BandData} now returns values suitable for the
selection, so that user classes of \texttt{BandDataBZ} cannot
distinguish whether the entire band or only a part is used,
allowing them to be treated uniformly. As shown before, this
loose-coupling between components helps modification and
extension of the program.

\subsection{\texttt{Properties}}
While in principle all calculations can be done only with the
data in the \texttt{BandData} class, some specific calculations
occur frequently, so we encapsulate them in classes derived
from \texttt{Properties} to allow data reuse and save time. The
\texttt{Properties} interface only contains abstract I/O
functions (\texttt{read} and \texttt{print}) and a \texttt{name} function (Fig.
\ref{fig:design:hierarchy}), with the intent that the required
properties are calculated together and the I/O of them are done
together in the main program. C++ uses streams to do I/O, and one can provide
custom extraction (\texttt{>>}) and insertion (\texttt{<<})
operators to use user-defined classes together with streams.
With the help of the abstract I/O functions of
\texttt{Properties}, the extraction and insertion operators can
be implemented by the following pseudo-code:
\begin{verbatim}
istream& operator>>(istream& is, Properties& prop)
{
 string tmp_name;

 is >> tmp_name;

 if(tmp_name != prop.name()) // wrong file
  throw runtime_error(/*error message*/);

 return prop.read(is);
}

ostream& operator<<(ostream& os,
 const Properties& prop)
{
 os << prop.name();
 return prop.print(os);
}
\end{verbatim}
We use C++ standard exception handling here, so
in the main program doing I/O we can recover from reading a
corrupted file:
\begin{verbatim}
try
{
 file >> prop;
}
catch(const runtime_error& err)
{
 if(err.what()==.../*I/O error message*/)
  .../*construct prop from beginning
     instead of from file*/
 else throw;
}
\end{verbatim}
This has the advantage over returning an error code, because
the exception cannot be ignored, and the exception handling can
happen far away from the actual point of error. In the previous
example, the error can happen both in \texttt{operator>>} and
in the \texttt{read} method of a derived class of
\texttt{Properties}. For a program using error codes, they need
to be propagated and checked manually at every step, while C++
exceptions propagate automatically and will terminate the
program if not handled. The resources are also automatically
freed with the help of smart pointers.\cite{ExceptionalCpp}
This not only makes the program more structured (by grouping
error handling together), but also automatically prevents the
program from being in an inconsistent state when an error
occurs.

\texttt{Properties} is meant to hold frequently used
intermediate results, such as
$\matelem{j\vect{k}}{e^{i\vect{G}\cdot\vect{r}}}{i\vect{k}}$ in
Eq. \parref{eqn:intro:fxcworking}, calculated by the
\texttt{Prop\_ExpIGr} class. Therefore it is crucial that these
results can be accessed efficiently. We heavily use the C++'s
standard template library (STL)\cite{STL}
as a way of organizing data, so that the time and
space cost for accessing the data is well-defined and easy to
change for different uses.

For example, \texttt{Prop\_ExpIGr} values are labeled by $i$,
$j$, $\vect{k}$, and $\vect{G}$. We define a custom
\texttt{expigr\_key} type to hold these values, and use it as
the key of the standard \texttt{map} container. The
\texttt{map} container guarantees fetching and inserting value
in logarithmic time. Its interface overloads the array access
operator (\texttt{operator[]}), so the use resembles the regular arrays. This
makes accessing the values both efficient and intuitive. If
later we need faster access of the results, we can change to
use the standard \texttt{unordered\_map} container which
guarantees constant time access. Only one line of the code
needs to be changed, and the implementation of
\texttt{Prop\_ExpIGr} does not need to change thanks to the
similar interface between all standard containers.

STL also benefits from OOP, in the sense that only the
interface is visible to the user. Different implementations of
STL are available: some optimize for speed, some for memory
usage, and some for thread safety. These implementations are
completely decoupled from the codes using them, so we can
change to a different implementation without changing anything
in our source code. Separating interface and implementation is
of course not a new concept, but OOP makes it easier and safer
to use.

We calculate properties beforehand to exploit parallelism. We
implement two level of parallelism in our code, the first being
process (an instance of the program with its own address space)
level implemented with the message-passing interface\cite{MPIstandard}
(MPI), and the second being thread level implemented with
OpenMP\cite{OpenMP} compiler directives. We presume that communication
between processes (which are running on
different nodes of a cluster) is slow, and thus we minimize the
number of communications and group communications together.

The derived classes of \texttt{Properties} involve doing the
same calculation on different sets of data. We implement
parallelism accordingly. We generate all the jobs that need to
be done and pass roughly the same number of jobs to each
process with MPI, and then use OpenMP to parallelize the
calculation of the jobs on each process. When all calculations
are done, we collect the results and distribute them to all
processes using MPI. OpenMP is a standard for specific compiler
directives, designed to be able to be added to a serial program
without changing its structure. For non-conforming compilers,
these compiler directives are ignored, and the program reverts
to a serial program automatically.

Unlike the OpenMP parallelism which can be easily added, the
MPI standard is designed with imperative programming in mind.
The MPI subroutines (even with its C++ binding) operate
on raw data. The user is required to consider low-level details
such as the buffer or the memory alignment, and this does not
work well together with OOP. The
\texttt{Boost::MPI}\cite{boost} library encapsulates the low-level MPI
subroutines in an object-oriented manner. With very few extra
programming, \texttt{Boost::MPI} allows objects of user-defined
types to be used in the same syntax as data of fundamental MPI
types, and the low-level details involved in using user-defined
types are determined automatically. The calculation of a
derived class of \texttt{Properties} is done by the following
pseudo-code:
\begin{verbatim}
mpi::communicator world;
vector<key> myjob;

if(myid == ID_Master)
{
 vector<key_type> jobs;

 // generate all the jobs on one node
 for( .../*all possible jobs*/ )
  jobs.push_back( .../*one job*/ )

 vector< vector<key_type> > splitted_jobs;

 /*prepare the jobs to be distributed to each
 nodes, store in `splitted_jobs' */
 ...

 /* distribute the jobs */
 mpi::scatter(world,splitted_jobs,myjob,ID_Master);
} // jobs and splitted_jobs are destroyed here
/* receive the jobs for this node */
else mpi::scatter(world,myjob,ID_Master);

size_t pos;
// OpenMP parallelization for jobs on this node
#pragma omp parallel for
for(pos = 0; pos < myjob.size(); ++pos)
 // store result in a container local_result
 DoCalculation(myjob[pos]);

result_type temp;

// gather results and distribute to all nodes
mpi::all_reduce(world, local_result, temp,
 .../*an function object combining results*/);

local_result.swap(temp);
// now local_result contains all results
\end{verbatim}
The C++ STL containers as well as user-defined types are used
directly with \texttt{Boost::MPI} without exposing the
underlying raw data. STL does not specify thread-safe write access,
but thread-safety can be easily achieved with OpenMP
synchronizing constructs such as the critical region. The
pseudo-code for \texttt{DoCalculation} appeared above is
\begin{verbatim}
void DoCalculation(const key_type& key)
{
 /* do the calculation indicated by key,
 store in variable `result' */
 ...

 // OpenMP critical region
 #pragma omp critical
 {
  local_result.insert(result);
 }
}
\end{verbatim}
This guarantees that only one thread writes to the container at
a time. Though critical regions are costly, in practice we find
that the actual calculation takes much more time, compared to
which the cost of the critical region is acceptable.

\subsection{\texttt{fxc}}
The xc kernel $f\xc$ is central to a TDDFT calculation. Many
approximations have been developed since the exact $f\xc$ is
unknown. In our recent study,\cite{YU13} different $f\xc$'s are
treated uniformly as in Eq. \parref{eqn:intro:fxcworking}, but
we use several $f\xc$'s which are different in their own
details: some only need trivial calculation, but some require
extra data and significant calculation. These makes $f\xc$
ideal to benefit from polymorphism.

The abstract \texttt{fxc} base class defines an interface (Fig.
\ref{fig:design:hierarchy}) for accessing the value of $f\xc$
in reciprocal space (via \texttt{fxcQspace} and related
methods) and for I/O. The I/O of \texttt{fxc} is done similarly
as in \texttt{Properties}. It should be noted that the
polymorphism of \texttt{fxc} can also be achieved by function
pointers in imperative programming. Instead of an object of
\texttt{fxc} type, the calculation of Eq.
\parref{eqn:intro:fxcworking} can require several function
pointers playing the role of \texttt{fxc::fxcQspace} and
related methods.

The function pointer approach works well for simple xc kernels
that can be calculated on-the-fly. However, xc kernels
requiring heavy calculation (such as the `bootstrap'
kernel\cite{SDSG11}) must be calculated beforehand and the
results must be stored; with function pointers this is only
achievable by having stronger coupling between different parts
of the program --- one has to find a place to store these intermediate data,
and thus making it harder to maintain. With the
object approach, however, intermediate results are conveniently
stored in private member variables and are concealed from other
parts of the program. As shown previously, the major difference between
OOP and imperative programming implementations is whether the
relation between data (pre-calculated xc kernel) and
subroutines manipulating it (\texttt{fxcQspace} etc.) can be
explicitly expressed through native constructs of the
programming language. This expressiveness of OOP compared to
imperative programming allows more error-checking to be
delegated to the compiler, reducing possible human error.

The polymorphism of \texttt{fxc} also helps when doing triplet
calculations. As shown in Sect. \ref{sect:intro:bkgnd},
singlet/triplet excitations for spin-unpolarized system can be
calculated by using different xc kernels as in Eq.
\parref{eqn:intro:spinfxc}. We derive the abstract
\texttt{fxc\_spin} class from \texttt{fxc}, which contains an
additional singlet/triplet mode switch. Since
\texttt{fxc\_spin} {\em is an} \texttt{fxc}, it is used in the
same way for evaluating Eq. \parref{eqn:intro:fxcworking}. One
obtains both singlet and triplet results by doing two
successive calculations with the corresponding mode of
\texttt{fxc\_spin}. The code calculating Eq.
\parref{eqn:intro:fxcworking} neither knows that the calculation is
spin-dependent, nor can a bug change the mode accidentally.

\subsection{\texttt{Hamiltonian}}
We define the \texttt{Hamiltonian} class to construct the
matrices in Eqs. \parref{eqn:intro:fHworking} and
\parref{eqn:intro:fxcworking} and to solve the eigenvalue
problem Eq. \parref{eqn:intro:TDDFTworking}. The storage and
I/O of data is one of the main problems. The matrices have
dimension $N_v\times N_c\times N_\vect{k}$, where $N_v$ is the
number of valence bands, $N_c$ is the number of conduction
bands, and $N_\vect{k}$ is the number of $\vect{k}$-points in
the full Brillouin zone. For a converged calculation, the
number of matrix elements must be at least of the order of
$10^8$. \texttt{Hamiltonian} thus contains hundreds of
gigabytes of data in total. Due to hardware limitations, we are
forced to use distributed storage and parallel algorithms to
handle these matrices.

\begin{figure}
\includegraphics[width=\columnwidth]{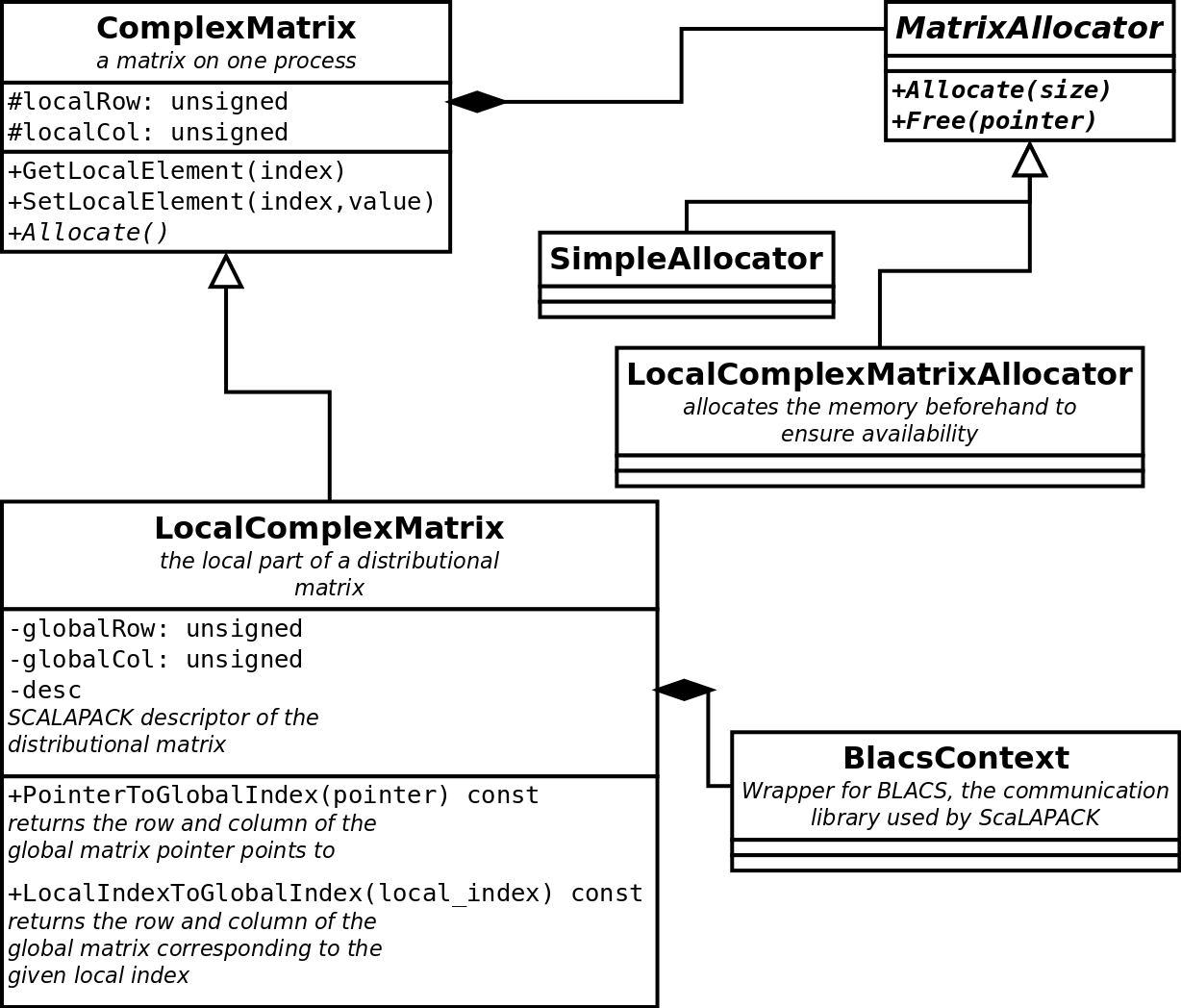}
\caption{UML\cite{UML} class diagram of \texttt{ComplexMatrix} and \texttt{LocalComplexMatrix}. Only important methods and member variables are shown.}
\label{fig:impl:complexmatrix}
\end{figure}

The ScaLAPACK library\cite{ScaLAPACK} nicely suits our needs.
It provides expertly tuned parallel algorithms for distributed
matrices, but its interface exposes many implementation details
that do not fit into the OOP framework. For example, the local
storage format of the distributed matrix must be provided when
calling most of the subroutines of ScaLAPACK. If
\texttt{Hamiltonian} had to manage such details, it would
create lots of duplicate code,  which would be hard to
maintain. Changing to other libraries would then involve major
rewriting of the \texttt{Hamiltonian} class. We package the use
of numerical libraries in the \texttt{ComplexMatrix} and
\texttt{LocalComplexMatrix} classes (Fig.
\ref{fig:impl:complexmatrix}).

We use \texttt{ComplexMatrix} to store a matrix on one process:
it allocates the necessary memory through a provided object of
the \texttt{MatrixAllocator} class, it gives access to the
matrix elements, and it provides commonly used operations such
as adding two matrices. By wrapping several BLAS\cite{BLAS} and
LAPACK\cite{LAPACK} subroutines for computation-heavy tasks,
\texttt{ComplexMatrix} not only provides numerical efficiency,
but also hides the representation of the matrix, making
substituting BLAS/LAPACK with other libraries possible.

For the use of matrices in the \texttt{Hamiltonian} class, we derive
\texttt{LocalComplexMatrix} from \texttt{ComplexMatrix} to
represent the local part of a distributed matrix. The code in
\texttt{ComplexMatrix} is thus reused for operations not
directly related to the distributed nature, such as adding two
matrices. To solve the eigenvalue problem Eq.
\parref{eqn:intro:TDDFTworking} represented in a
\texttt{LocalComplexMatrix} object, we define an abstract
\texttt{EigenSolver} class, whose derived classes package the
actual ScaLAPACK subroutine calls. By doing so the
\texttt{Hamiltonian} class is decoupled from the representation
details of the distributed matrix.

The I/O of the distributed matrices is problematic, because
simply saving the data on each node to disk forbids using a different
number of nodes when reusing this data. We use MPI to deal with
the I/O problem. ScaLAPACK uses a block-cyclic format for
distributed matrices, which is supported by the MPI 2.0
standard as a representable data type. The MPI parallel I/O
interface also contains unrelated details for the
\texttt{Hamiltonian} class, so we package them into the
\texttt{HamilInOutAdapter} class. In the end, the
\texttt{Hamiltonian} class only determines what calculations
need to be done, but does not depend on how they are actually
carried out. In this way, the code achieves a high degree of
flexibility and extensibility.



\section{Conclusion}
\label{sect:conclusion}

We have presented and examined our code for the recent study of
exciton binding energies with TDDFT.\cite{YU13} We provide our
code as a case-study showing the advantages of OOP and
promoting the use of OOP in scientific programming. The source
code is made available as supplemental
material,\cite{supplemental}  but it is not intended to be used
as a black box; it only serves as an example to illustrate the
power of OOP and to provide the background of this work.

Compared with imperative programming, OOP helps to analyze and
model the computational problem at a level closer to the actual
physical problem. By providing an explicit connection between
data and operations manipulating the data (through programming
language support), the structure of the program becomes
loosely-coupled, i.e., changing one part of the code does not
affect the behavior of other parts. Consequently the
maintenance of the program becomes easier.

The versatility and expressiveness of OOP allows delegating
many types of error-checking to the compiler, allowing earlier
detection of bugs and reduction of human error. Encapsulating
irrelevant details inside objects frees physicists from being
entangled by implementation details and instead allows
concentrating on the physical problem at hand. Like any other
programming paradigms, OOP itself does not guarantee a good
program, but the abstraction provided by it helps achieving
higher programming quality.

\section*{Acknowledgement}
This work is funded by the National Science Foundation Grant No. DMR-1005651.

\bibliography{draft}
\bibliographystyle{unsrt}

\end{document}